\documentclass[aps,prb,twocolumn,floatfix,10pt] {revtex4-2}
\usepackage{graphicx,amsfonts}
\usepackage{bm,color}
\usepackage{multirow}
\usepackage{amssymb,amsmath,hyperref}
\usepackage{wrapfig}
\usepackage{float}
\DeclareUnicodeCharacter{2212}{-}
\usepackage[table]{xcolor}

\begin{document}

%\title{Driven flux qubit strongly coupled to a quantum resonator}
\title{Probing strongly driven and strongly coupled superconducting qubit-resonator system}

\author{Oleh V. Ivakhnenko$^{1,2}$}\email{oleh.ivakhnenko@riken.jp}
\altaffiliation{Contributed equally to this work}

\author{Christoforus Dimas Satrya$^{3}$}\email{christoforus.satrya@aalto.fi}
\altaffiliation{Contributed equally to this work}

\author{Yu-Cheng~Chang$^3$}

\author{Rishabh Upadhyay$^3$}

\author{Joonas T. Peltonen$^3$}

\author{Sergey N. Shevchenko$^1$} %\email{sshevchenko@yahoo.com}

\author{Franco Nori$^{2,4}$}

\author{Jukka P. Pekola$^3$}

\affiliation{$^1$ B. Verkin Institute for Low Temperature Physics and Engineering, Kharkiv 61103, Ukraine}

\affiliation{$^2$ Center for Quantum Computing, RIKEN, Wako, Saitama, 351-0198, Japan}

\affiliation{$^3$ Pico group, QTF Centre of Excellence, Department of Applied Physics, Aalto University, P.O. Box 15100, FI-00076 Aalto, Finland}

\affiliation{$^4$ Physics Department, The University of Michigan, Ann Arbor, MI 48109-1040, USA}

\date{\today}

\begin{abstract}

We investigated a strongly driven qubit strongly connected to a quantum resonator.
The measured system was a superconducting flux qubit coupled to a coplanar-waveguide resonator which is weakly coupled to a probing feedline. This hybrid qubit-resonator system was driven by a magnetic flux and probed with a weak probe signal through the feedline. We observed and theoretically described the quantum interference effects, deviating from the usual single-qubit Landau-Zener-St\"{u}ckelberg-Majorana interferometry, because the strong coupling distorts the qubit energy levels.

\end{abstract}

\maketitle

\section{Introduction}

Circuit quantum electrodynamics provides a platform for quantum phenomena experiments and quantum technology applications %blais_2020_quantum
\cite{Xiang2013,Gu2017,Krantz2019,Kjaergaard2020,clerk_2020_hybrid,carusotto_2020_photonic,blais_2021_circuit}. Natural microscopic quantum systems, when driven or coupled, experience this perturbation usually weakly \cite{Gustafsson2014}. In contrast, artificial quantum systems allow to realize both strong coupling and strong driving \cite{Oliver2005,Oliver2009,Wen2018}. The layout when natural or artificial atoms are \textit{strongly coupled} to
cavities provides both new phenomena in quantum physics and is important for
quantum engineering. Strong coupling \cite{Ashhab2010,Hu2012,Flick2018, Dovzhenko2018,FriskKockum2019,FornDiaz2019,Mercurio2022,Bjoerkman2025} was recently studied for both natural 
 and artificial atoms. In
quantum computation, strong coupling increases the speed of gate operations
and information exchange \cite{PhysRevLett.130.233602}.

For the control of quantum systems \textit{ strong periodic driving}
provides a convenient tool. Since theoretically this relates to the
transitions between energy levels and their interference, this regime
can be studied using Landau-Zener-St\"{u}ckelberg-Majorana (LZSM)
interferometry \cite{Sillanpaeae2006,Ashhab2007,You2007,Shevchenko2010,Shevchenko2012,Silveri2017,Ivakhnenko2023,Ryzhov2024,Kivelae2024,Tedo2025,Peyruchat2025,Hu2025,Bjoerkman2025}.

Here we consider two important regimes of operation for
quantum system: strong coupling and strong driving, both of which
have attracted attention. However, these two aspects
rarely meet and are realized in one device. Recently efforts in this
direction were done in Ref.~\cite{PhysRevLett.130.233602}, which studied two
strongly driven double quantum dots strongly coupled to a cavity.%

In this work, we study a flux qubit strongly coupled to a resonator. We measure the transmission of the driven hybrid qubit-resonator system and observe quantum interference effects. We are able to reproduce the measured interference with our theoretical model. This system provides a useful framework for quantum thermodynamic experiments, for example for measuring heat produced by a driven qubit \cite{pekola_2013_calorimetric,Yan2016,PhysRevResearch.5.L022036,Zhelnin2025}, and the realization of a quantum heat engine and refrigerator, where a driven qubit is coupled to two dissipative resonators \cite{PhysRevB.94.184503,PhysRevE.102.030102}, and a heat engine/refrigerator cycles for a single qubit with modulated energy levels \cite{Ono2020,Guthrie2022,uusnäkki2025experimentalrealizationquantumheat}.

\begin{figure*}[ht]
\includegraphics{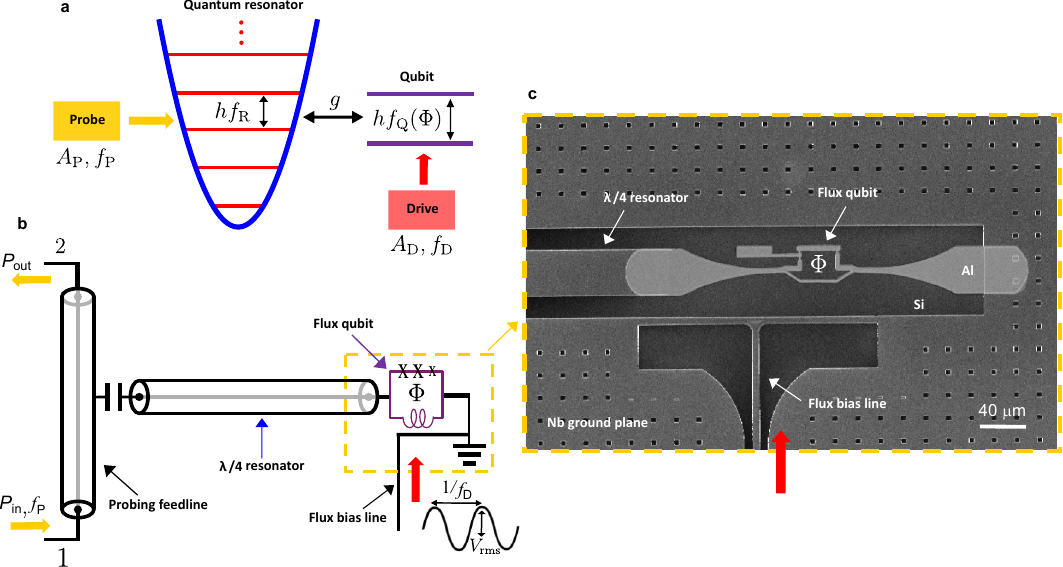} \caption{\textbf{a}, A quantum resonator is coupled strongly to a qubit via $g$. The hybrid system
is driven by two signals, probe and drive tones, with frequencies $f_{\textrm{P}}$ and $f_{\textrm{D}}$, respectively, and their amplitudes $A_{\textrm{P}}$ and $A_{\textrm{D}}$. \textbf{b}, The physical realization of the qubit-resonator device. A $\lambda/4$ resonator is capacitively coupled to a probing feedline and shorted by a flux qubit with three Josephson junctions in parallel with an inductor $L$ \cite{upadhyay_2021_robust}. \textbf{c}, Scanning electron microscope image of the fabricated device shows the resonator shunted by the flux qubit. The on-chip flux bias line is located nearby the qubit to inject the drive tone. The resonator and feedline are made of 200 nm-thick Nb film (bright gray) on top of AlO$_x$/Si substrate (dark gray), while the flux qubit is made of Al film (light grey). The transmission $|S_{21}|$ measurement is carried out through the feedline of Port-1 and Port-2. See Fig.~\ref{fig:8} for the whole layout of the device.\label{fig:1}}   
\end{figure*}

\section{The measured system and device}

A quantum resonator is coupled to a qubit, with a coupling $g$, as depicted in Fig.~\ref{fig:1}(a). The characteristic energy of the resonator is $hf_{\textrm{R}}$ and the qubit energy is $hf_{\textrm{Q}}$ that can be tuned by a magnetic flux $\Phi$. The hybrid qubit-resonator system is driven by two signals. The first one is a probe tone with amplitude $A_{\textrm{P}}$ and frequency $f_{\textrm{P}}$, which excites and probes the resonator. The second signal is a drive tone, with amplitude $A_{\textrm{D}}$ and frequency $f_{\textrm{D}}$, which modulates cyclically the qubit energy.

The physical realization of this system is shown in Fig.~\ref{fig:1}(b). A quarter-wavelength ($\lambda/4$) coplanar waveguide (CPW) resonator is capacitively connected to a probing feedline. The fundamental-mode resonance of the resonator is $f_{\textrm{R1}}$, with the higher-mode resonances: $f_{\textrm{R3}}=3f_{\textrm{R1}}$ and $f_{\textrm{R5}}=5f_{\textrm{R1}}$.

Here we focus on the interaction between the qubit and the mode resonance $f_{\textrm{R3}}$, which we define as the resonance frequency $f_{\textrm{0}}$. The shorted end of the resonator is shunted galvanically by a flux qubit composed of a parallel inductor $L$ and three Josephson junctions \cite{upadhyay_2021_robust} where two of them are designed to be identical in size, shown by the scanning electron microscope (SEM) image in Fig.~\ref{fig:1}(c). The strength of the interaction between resonator and qubit can be controlled by the inductance $L$, $g\sim L$. The flux qubit Hamiltonian is
\begin{equation}
    H_{\textrm{Q}} = -\frac{1}{2}\hbar(\Delta\sigma_{x} + \varepsilon\sigma_{z}),
    \label{eq:han_def}
\end{equation} where $\hbar\varepsilon = 2I_{\mathrm{p}}(\Phi - \Phi_0/2)$,  $I_{\mathrm{p}}$ stands for the persistent current, $\Delta$ is the minimal qubit energy splitting, $\Phi_0$ is the magnetic-flux quantum, and $\Phi$ is the external magnetic flux applied through the qubit loop. The qubit energy is

\begin{equation}
 \omega_\text{Q}=2\pi f_{\textrm{Q}} =\sqrt{\Delta^2+\varepsilon^2},
    \label{eq:h_def}
\end{equation}where $\varepsilon$ is tuned by applying the flux $\Phi$ through the loop. In the experiments, we can apply static (DC) and alternating (AC) fluxes to the qubit loop as

\begin{equation}\label{eq:ElectrPhonon}
  \Phi = \Phi_{\textrm{DC}} + \Phi_{\textrm{AC}}\sin(2\pi f_{\textrm{D}}t),
\end{equation}where the second term originates from the drive tone. The DC flux ($\Phi_{\textrm{DC}}$) is controlled by applying a DC current to the coil located outside the sample holder. The AC flux is controlled by the drive tone injected through an on-chip flux bias line inductivelly coupled to the qubit loop. The drive tone is generated by applying a sinusoidal voltage from a signal generator with voltage rms $V_{\textrm{rms}}$ and frequency $f_{\textrm{D}}$.

%%%The amplitude $V_{\textrm{rms}}$ determines $\Phi_{\textrm{AC}}$ and drive amplitude $A_{\textrm{D}}$.  

\begin{figure}[H]
	\includegraphics{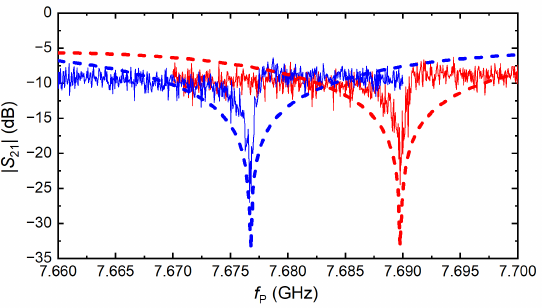}
	\caption{Measured transmission $|S_{21}|$ (with $P_{\textrm{in}}=-40~\textrm{dBm}$ and $A_{\textrm{D}}=0$) is plotted using solid curves. The blue fluctuating curve is at a flux bias $\Phi_{\textrm{DC}}/\Phi_{\textrm{0}}=0$, with the resonator resonant frequency $f_0=7.6767~{\textrm{GHz}}$. The total photon relaxation rate, that defines the line-width is approximately equivalent to $\kappa/2\pi\sim4.71~\textrm{MHz}$. The red fluctuating curve is at a flux bias $\Phi_{\textrm{DC}}/\Phi_{\textrm{0}}=0.5$, with resonant frequency of the qubit-resonator system at $f_\text{QR}=7.6898~{\textrm{GHz}}$. Thus, the dispersive shift is $\chi=13.1~\textrm{MHz}$. The dashed curves are the theoretical transmissions, numerically calculated from the Lindblad equation with Hamiltonian~\eqref{DressedRWAHamiltonian} and transmission coefficient~\eqref{PhotonNumberAv}.\label{fig:2}}   
\end{figure}

\begin{figure*}[ht]
	\includegraphics{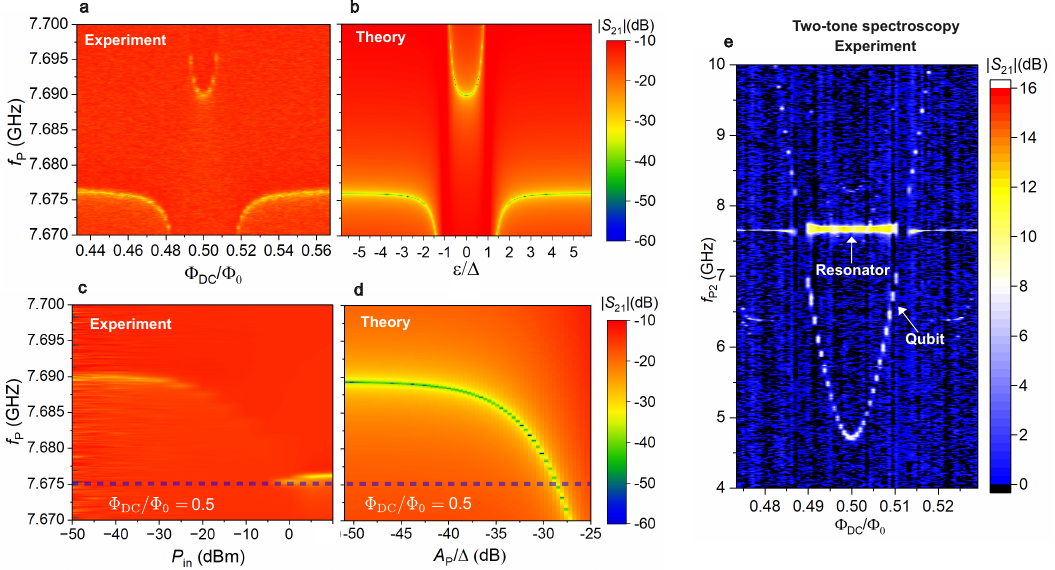}
	\caption{Spectroscopy of the qubit-resonator system. Here the system is driven only by the probe tone through the resonator from the feedline with $P_{\textrm{in}}=-40~\textrm{dBm}$. The qubit drive from the flux-bias line is off, $A_{\textrm{D}}=0$. One-tone spectrosopy: \textbf{a}, experiment and \textbf{b}, theory. \textbf{c-d}, Spectroscopy at fixed flux bias $\Phi_{\textrm{DC}}/\Phi_{\textrm{0}}=0.5$ with varying probe amplitude: \textbf{c}, experiment and \textbf{d}, theory. The horizontal dashed purple lines indicate the resonance frequency of the resonator $f_0$. \textbf{e}, Two-tone spectroscopy experiment showing the interaction of the qubit and resonator energy levels. \label{fig:3}}   
\end{figure*}

A probe signal with power 
$P_{\textrm{in}}$ and frequency $f_{\textrm{P}}$ is applied through Port-1 of the feedline, and the output signal coming out from Port-2 of the feedline is measured. With a vector network analyzer (VNA), the scattering parameter (transmission), whose amplitude is
\begin{equation}
	|S_{21}(f)| = \sqrt{P_{\textrm{out}}(f)/P_{\textrm{in}}(f)},
\end{equation} can be measured. The power $P_{\textrm{in}}$ determines the probe amplitude $A_{\textrm{P}}$. The detailed measurement setup and fabrication of the device are described in Appendices \ref{Fabrication_methods} and~\ref{Experimental_results}.

\section{Theoretical description}

\subsection{Hamiltonian}
\label{Sec:Introduction}

 Here we describe the system theoretically with numerical simulation of the modified Jaynes–Cummings Hamiltonian to investigate probing and driving of the qubit-resonator system. First, we write the Hamiltonian of the flux qubit-resonator system in the diabatic basis \cite{Omelyanchouk2010,Guthrie2022,sahelashhab_2023_controlling},
\begin{subequations} 
\begin{eqnarray}
H_\text{tot}=H_\text{Q}+H_\text{R}+H_\text{C}+H_\text{P}+H_\text{D},\label{InHam} \\
	H_\text{Q}=\hbar\left(-\frac{\Delta}{2}\sigma_x-\frac{\varepsilon_0}{2}\sigma_z\right),\\
	H_\text{R}=\hbar\omega_\text{r1}\left(a^\dagger a+\frac{1}{2}\right),\\
	H_\text{C}=-\hbar g\left(a+a^\dagger\right)\sigma_z,\\
	H_\text{P}=\hbar A_\text{P}\left(ae^{i\omega_\text{p}t}+a^\dagger e^{-i\omega_\text{P}t}\right),\\
	H_\text{D}=-\hbar\frac{A_\text{D}\sin \omega_\text{D} t}{2}\sigma_z,
\end{eqnarray}
\end{subequations} 
where $H_\text{tot}$ is the total Hamiltonian of the system, $H_\text{Q}$ is the qubit Hamiltonian, with $\Delta$ being the minimal energy gap, $\varepsilon_0$ is constant detuning, and $\sigma_{x,y,z}$ are Pauli matrices. 
Moreover, $H_\text{R}$ is the resonator Hamiltonian, where $a$ and $a^\dagger$ are the annihilation and creation operators for the resonator, and $\omega_\text{r1}=2\pi f_{\textrm{0}}$ is the resonator frequency, which interacts with the qubit. 
$H_\text{C}$ is the coupling Hamiltonian, where $g$ is qubit-resonator coupling strength.
$H_\text{P}$ is the probe signal Hamiltonian, where the probe frequency for the interferometry is  $\omega_\text{P}\approx\omega_\text{r1}$ and the probe amplitude is $A_\text{P}$.
 $H_\text{D}$ is the qubit driving Hamiltonian, where $A_\text{D}$ is the driving amplitude, and $f_{\textrm{D}}=\omega_\text{D}/2\pi $ is the driving frequency.

\begin{table}[b]  
	\caption{Parameters used for the simulations} % title name of the table  
	\centering % centering table  
	\begin{tabular}{l c c c c} % creating 10 columns  
		\hline\hline   
		Quantity &Symbol &Value
		\\ [0.5ex]  
		\hline   
		% Entering 1st row  
& & \\[-1ex]  
		{Minimum qubit frequency} & {$\Delta/2\pi$}&$5.41~\mathrm{GHz}$
		\\[1ex]  
  & & \\[-1ex]  
		{Resonator first-mode frequency} & {$f_0/2\pi$}&$7.6767~\mathrm{GHz}$
		\\[1ex]  
  & & \\[-1ex]  
		{Qubit-resonator coupling} & {$g/2\pi$}&$177~\mathrm{MHz}$
		\\[1ex]  
  & & \\[-1ex]  
		{Number of photons in resonator} & {$N$}&$2$
		\\[1ex]  
   & & \\[-1ex]  
		{Relaxation rate} & {$\Gamma_1/2\pi$}&$3~\mathrm{MHz}$
		\\[1ex] 
   & & \\[-1ex]  
		{Dephasing rate} & {$\Gamma_2/2\pi$}&$1.5~\mathrm{MHz}$
		\\[1ex] 
   & & \\[-1ex]  
		{Resonator relaxation rate} & {$\kappa/2\pi$}&$4.71~\mathrm{MHz}$
		\\[1ex] 
    & & \\[-1ex]  
		{Probe amplitude} & {$A_{\mathrm{P}}/2\pi$}&$1~\mathrm{MHz}$
		\\[1ex] 
		% Entering 2nd row  
		% [1ex] adds vertical space  
		\hline % inserts single-line  
	\end{tabular}  \label{tab:1}
\end{table}

\begin{figure*}[ht]
\begin{center}  
\includegraphics{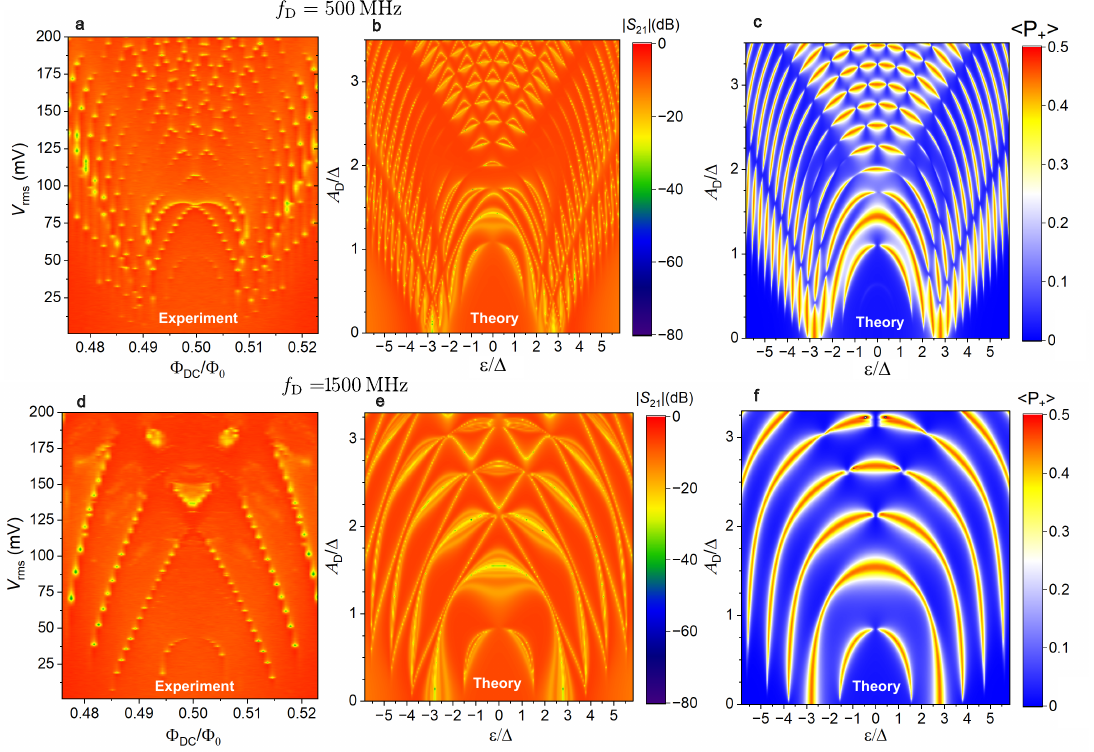}

\caption{Interferometry of the driven qubit-resonator system with fixed power $P_{\textrm{in}}=-40~\text{dBm}$ and driving frequency $f_{\textrm{D}}=500~\textrm{MHz}$  for the upper panels and $f_{\textrm{D}}=1500~\textrm{MHz}$ for the lower ones. The probe frequency $f_{\textrm{P}}$ is fixed at the resonance frequency $f_{0}$, and the drive amplitude is varied. Transmission: \textbf{a,d}, experiment and \textbf{b,e}, theory. \textbf{c,f}, The upper qubit population. \label{fig:4}}   
\end{center}

\end{figure*}

To study theoretically the transmission coefficient of the transmission line, we consider the absolute value of the imaginary part of the photon annihilation operator in the resonator \cite{Omelyanchouk2010}
\begin{equation}
	\left\langle |\text{Im}{(a)}|\right\rangle=\text{Tr}(|\text{Im}{(a)}|\rho)=|S_{21}|,
	\label{PhotonNumberAv}
\end{equation} 
where $\rho$ is the density matrix of the qubit-resonator system. In the limit of a weak probe signal, we can take into account only the first two-photon states in the resonator. After applying the rotating-wave approximation (RWA) and transfer to the instantaneous eigenstate basis, we obtain the dressed Hamiltonian in RWA
\begin{eqnarray}
	\label{DressedRWAHamiltonian}	
	&H_\text{RWA}=\hbar\left[\frac{1}{2}\delta\omega_\text{Q}\sigma_z+\delta\omega_{r1}a^\dagger a\right. \\&\left. +g\frac{\Delta}{\omega_\text{Q}} \left(\sigma_+a+\sigma_-a^\dagger\right)+A_\text{P}\left(a+a^\dagger\right)\right]\notag,
\end{eqnarray}
where 
\begin{subequations} 
\begin{eqnarray}
	\delta\omega_\text{Q}=\omega_\text{Q}-\omega_\text{P}, \\
	\omega_\text{Q}=\sqrt{\Delta^2+\varepsilon(t)^2},\\
	 \delta\omega_{r1}=\omega_{r1}-\omega_\text{P},\\
	\delta\omega_\text{D}=\omega_\text{D}-\omega_\text{P}.
\end{eqnarray}
\end{subequations} 
Here $\omega_\text{Q}$ is the instant qubit energy, as in Eq.~\eqref{eq:h_def}.
We now use this dressed RWA Hamiltonian to describe the theoretical spectroscopy of our qubit-resonator system for both the wide and narrow ranges of the probing frequency $\omega_\text{P}$ in Fig.~\ref{fig:3} and interferometry in Fig.~\ref{fig:4}. We use the Hamiltonian of Eq.~\eqref{DressedRWAHamiltonian} in the Lindblad equation~\eqref{LindbladEquation} with Lindblad operators~\eqref{Lindblad_operators} for the numerical calculations, more details of the theoretical description and transfer between bases and dressed states are presented in Appendix~\ref{Appendix_Theory}.
To numerically simulate this system, we solved the Lindblad master equation using the QuTiP library \cite{Johansson2012,Johansson2013,Lambert2024}.

\begin{figure*}[t]
	\begin{center}  
		\includegraphics{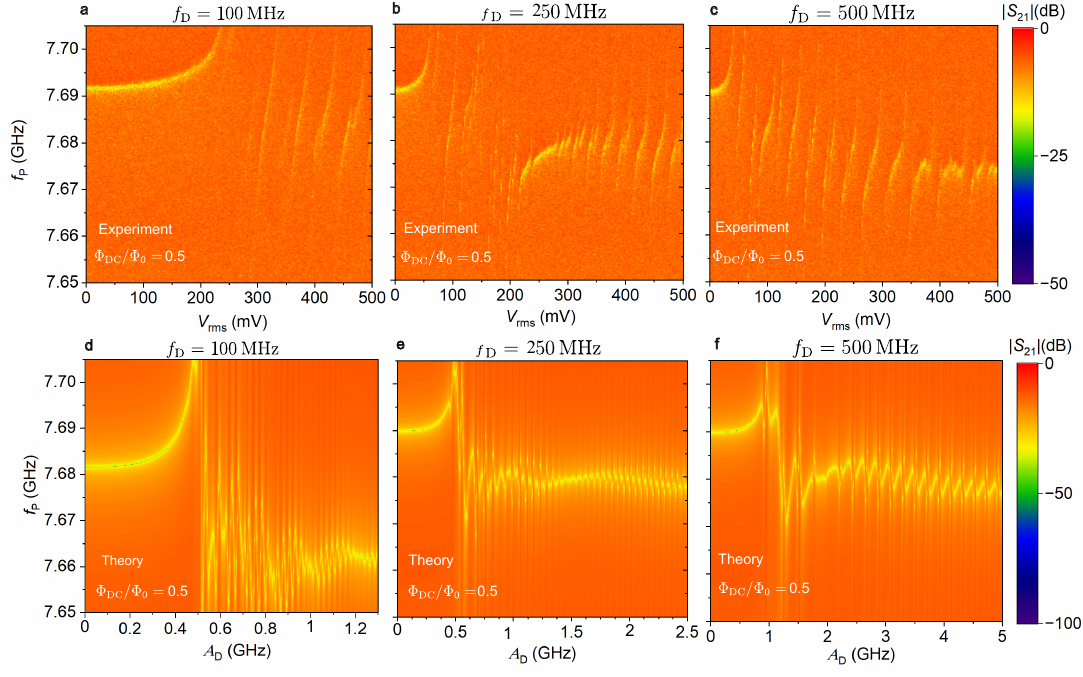}
		
		\caption{Interferometry of the driven qubit-resonator system with fixed power $P_{\textrm{in}}=-40~\text{dBm}$ and at fixed flux bias $\Phi_\text{DC}/\Phi_0=0.5$. The probe frequency $f_{\textrm{P}}$ and drive amplitude $A_{\textrm{D}}$ are varied. \textbf{(a-c)} are the measurement results at different drive frequency $f_{\textrm{D}}$, and \textbf{(d-f)} are the simulation results.\label{fig:5}}   
	\end{center} 
\end{figure*}

\section{Results and Discussion
}
The transmission ($S_{21}$) and spectroscopy of the device are presented in Fig.~\ref{fig:2} and Fig.~\ref{fig:3}, where the system is driven only by the probe tone through the resonator while the drive tone from the flux-bias line is off. The transmission of the experimental data is subtracted to reach the same level of background as in the simulation. Transmission at $\Phi_{\textrm{DC}}/\Phi_{\textrm{0}}=0$, where the resonator is fully decoupled from the qubit, shows the resonance of the resonator $f_{\textrm{0}}$, shown in Fig.~\ref{fig:2}. At $\Phi_{\textrm{DC}}/\Phi_{\textrm{0}}=0.5$ the measured resonance is pushed toward higher frequency due to dispersive interaction with the qubit. Sweeping the $\Phi_{\textrm{DC}}/\Phi_{\textrm{0}}$ towards 0.5 shifts the measured resonance frequency $f_\text{QR}$, due to the hybridization of the resonator and qubit.

When the qubit energy is in resonance with $f_{0}$, the Rabi splitting is observed, as shown in Fig.~\ref{fig:3}(a). The coupling $g$ is observed to be around $177~\mathrm{MHz}$. We calculate the transmission theoretically, and by adjusting the parameters $g$ and $\Delta$ we achieve good agreement between theory and the experiment, as shown in Fig.~\ref{fig:3}(b). 

We also fix the flux bias at $\Phi_{\textrm{DC}}/\Phi_{\textrm{0}}=0.5$ and sweep the power $P_{\textrm{in}}$. At higher $P_{\textrm{in}}$, the qubit is overpopulated by the probe tone; thus the resonance shifts and reaches back to the frequency $f_{0}$ as shown in Fig.~\ref{fig:3}(c), which we can reproduce theoretically in Fig.~\ref{fig:3}(d). The purple dashed line shows the resonance line $f_{0}$. 

Moreover, a two-tone spectroscopy %\cite{Satanin2012} 
is performed to measure the qubit transitions and their interaction with the resonator as shown in Fig.~\ref{fig:3}(e). The second probe tone with frequency $f_{\textrm{P,2}}$ is applied through the feedline to excite the system. When $f_{\textrm{P,2}}$ is in resonance with the frequency of the dressed qubit state which is very close to qubit frequency at the large detuning, the $S_{\textrm{21}}$ signal of the first probe changes. It shows that the qubit energy folows the parabola of Eq.~\eqref{eq:h_def} and crosses the resonator energy at frequency $\sim7.6767~\mathrm{GHz}$. The presented two-tone spectroscopy data here is measured from the sister sample, where it is observed $\Delta\sim4.9~\mathrm{GHz}$.

Furthermore, the system is also driven by both the probe and drive signals. The probe frequency $f_{\textrm{P}}$ is fixed at $f_{\textrm{0}}$ and the drive frequency $f_{\textrm{D}}$ at either 500 MHz or 1500 MHz. The transmission coefficient $S_{21}$ is measured for varying $V_{\textrm{rms}}$ and sweeping $\Phi_{\textrm{DC}}/\Phi_{\textrm{0}}$ around 0.5. We apply, with a weak power, $P_{\textrm{in}}=-40 \textrm{ dBm}$ to probe the system. The results of the interferometry  for a two different driving frequencies $f_{\textrm{D}}=500~\textrm{MHz}$ and $f_{\textrm{D}}=1500~\textrm{MHz}$ shown in Fig.~\ref{fig:4}. The calculated upper qubit population is shown in Fig.~\ref{fig:4}(c,f). From these interferograms, we can see that the resonance lines density is the same for the transmission and qubit level occupation, but the shape of the resonances is different due to qubit-resonator quantum coupling.  Moreover, we fix the flux bias at $\Phi_{\textrm{DC}}/\Phi_{\textrm{0}}=0.5$ and sweep the probe frequency $f_{\textrm{P}}$ and drive amplitude $A_{\textrm{D}}$. The results of the interferometry are shown in Fig.~\ref{fig:5} (experiments and simulations) with several different drive frequencies $f_{\textrm{D}}$  shows similar behaviour in experiment and theory. The parameters used in the simulations are summarized in Table~\ref{tab:1}. The parameters $f_0, g$ and $\kappa$ are obtained from the measurements, while the other parameters are adjusted to get the best match between experiments and simulations.

\begin{figure*}[ht]
\begin{center}  
\includegraphics[width=\textwidth]{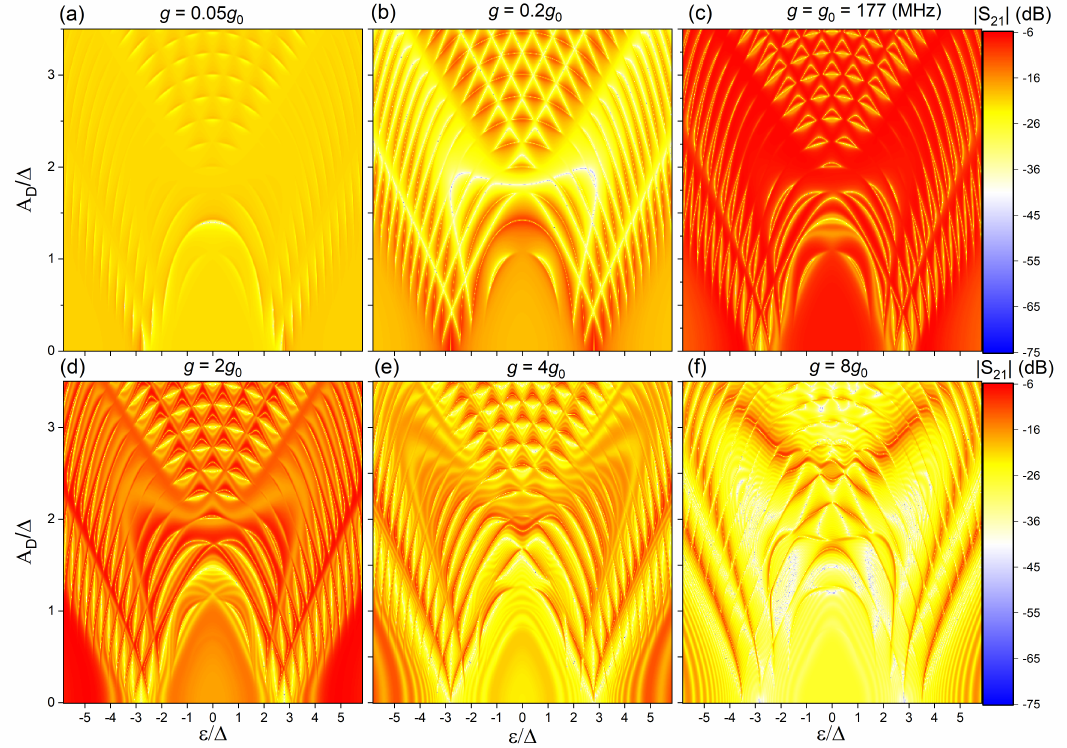}

\caption{Theoretical interferometry of the driven qubit-resonator system with fixed probing amplitude $A_\text{P}=1\text{ MHz}$ and driving frequency $\omega_{\textrm{D}}/2\pi=500\text{ MHz}$, for different strengths of the qubit-resonator coupling $g$, where $g_0=177\text{ MHz}$ was obtained from the spectroscopic calibration between theory and experiment shown in Fig.~\ref{fig:3}(a,b). The theoretical transmission coefficient $|\text{S}_{21}|$ is proportional to the averaged absolute value of imaginary part of the annihilation operator $|\text{Im}(<a>)|$. 
\label{fig:6}}   
\end{center} 

\end{figure*}

\begin{figure*}[ht]
\begin{center}  
\includegraphics[width=\textwidth]{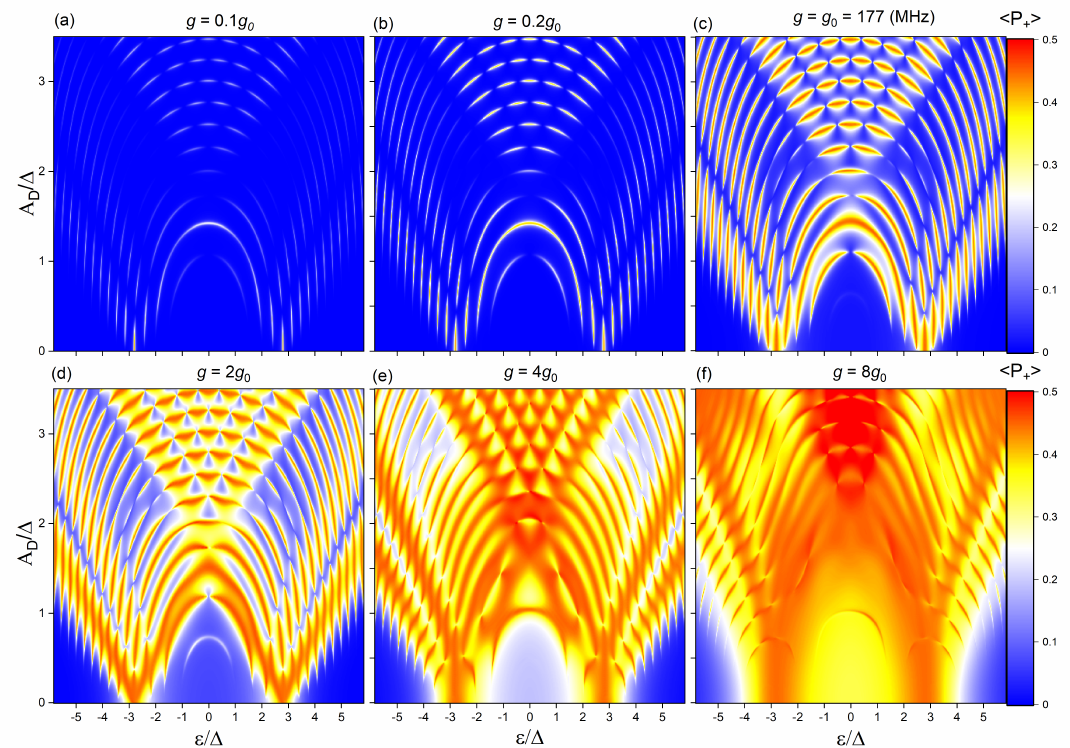}

\caption{Theoretical interferometry of the driven qubit-resonator system  with fixed probing amplitude $A_\text{P}=1\text{ MHz}$ and driving frequency $\omega_{\textrm{D}}/2\pi=500\text{ MHz}$, for different strengths of the qubit-resonator coupling $g$. For the upper level qubit occupation $P_+$ in the linear scale. Here $g_0=177\text{ MHz}$ was obtained from the spectroscopy calibration between theory and experiment, as shown in Fig.~\ref{fig:3}(a,b). 
\label{fig:7} }   
\end{center} 

\end{figure*}

\section{Conclusions
}

%We measure the transmission of driven strongly coupled resonator-flux qubit, and observe quantum interferences effects, different from the usual Landau-Zener-Stuckelberg-Majorana interferometry due to strong coupling between qubit and resonator. We notice that the biggest diefference from usual multiphoton LZSM interference appear at low driving field amplitudes, where impact of the strong coupling is more significant than impact of driving.
%Instant adiabatic basis, which take into account driving to the eigenstate basis transfer, is well suited for this problem and give good agreement with the experiment.
%Spectroscopy of the qubit is used for precise calibrating theoretical parameters of the system.
%With growing strength of coupling distortions of resonances increasing.
%Developed theory can be used to describe behavior of flux qubit strongly coupled to the resonator that can be used as a basis of the quantum heat engine.

We measured and calculated the transmission of a strongly driven coupled resonator-flux qubit system. We observed quantum interference effects, different from the usual Landau-Zener-St\"{u}ckelberg-Majorana interferometry due to strong coupling between the qubit and the resonator. The main difference from the usual multiphoton LZSM interference appears at low driving field amplitudes, where the impact of the strong coupling is more significant than the impact of driving. The instant adiabatic basis, which takes into account the driving to the eigenstate basis transfer, is well suited for this problem and gives good agreement with the experiment. The spectroscopy of the qubit is used for precisely calibrating the theoretical parameters of the system. When increasing the coupling strength, the distortions of the resonances become more pronounced. Our theory work here can be used to describe the behavior of a qubit strongly coupled to a quantum resonator, which could be used as a basis for a quantum heat engine and refrigerator in this regime where the resonator is more strongly coupled to the qubit than to the environment.

\section{Acknowledgments
}

\begin{figure*}[ht]
\includegraphics{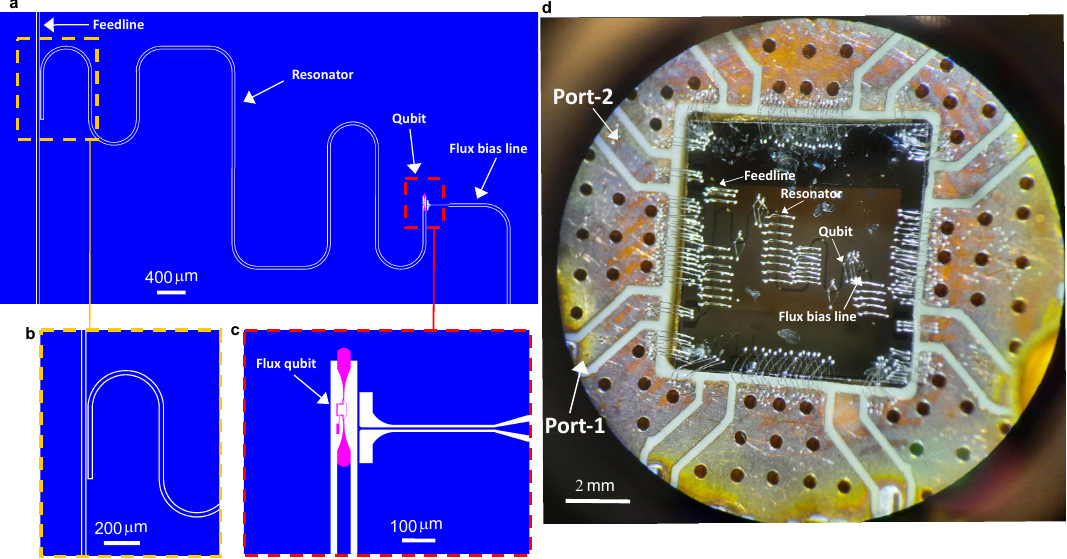} \caption{\textbf{a}, Layout of the fabricated device. The resonator is capacitevely coupled to the nearby feedline. The flux qubit shunts the other side of the resonator with a nearby flux bias line to drive the qubit. The blue background color represents the Nb film and the purple color represents the Al film. \textbf{b}, The feedline coupled to the resonator. \textbf{c}, The flux qubit coupled to the nearby flux bias line. \textbf{d}, The fabricated device with size 7mm x 7mm on top of the holder. The transmission $|S_{21}|$ measurement is carried out through the feedline of Port-1 and Port-2, shown on the left side of the disk. \label{fig:8}}   
\end{figure*}

This work is financially supported by the Research Council of Finland Centre of Excellence programme grant 336810 and grant 349601 (THEPOW). We sincerely acknowledge the facilities and technical supports of Otaniemi Research Infrastructure for Micro and Nanotechnologies (OtaNano) to perform this research. We thank VTT Technical Research Center for sputtered Nb films.
F.N. is supported in part by:
the Japan Science and Technology Agency (JST)
[via the CREST Quantum Frontiers program Grant No. JPMJCR24I2,
the Quantum Leap Flagship Program (Q-LEAP), and the Moonshot R\&D Grant Number JPMJMS2061],
and the Office of Naval Research (ONR) Global (via Grant No. N62909-23-1-2074), S.N.S. was partly supported by the Offiice of Naval Research (ONR)  and US National Academy of Sciences (NAS) IMPRESS-U grant  via STCU project~\#7120.

\appendix

\section{Detailed theoretical description}
\label{Appendix_Theory}
\subsection{Dissipative environment}
We use the Lindblad equation with initial Hamiltonian \eqref{InHam} to which we add relaxation and dephasing for the qubit and relaxation for photon states in the resonator
\begin{subequations} 
\begin{eqnarray}
	\label{Lindblad_operators}
	L_1^\text{D}=\sigma_-\sqrt{\Gamma_1},\\%S(\varepsilon=\varepsilon_0)\sqrt{(\Gamma_1)}\sigma_-S(\varepsilon=\varepsilon_0),\\
	L_2^\text{D}=\sigma_z\sqrt{\Gamma_2},\\%S(\varepsilon=\varepsilon_0)\sqrt{(\Gamma_2)}\sigma_zS(\varepsilon=\varepsilon_0),
	L_3^\text{D}=a\sqrt{\kappa},
\end{eqnarray}
\end{subequations} 
where $\Gamma_{1,2}$ are the relaxation and dephasing rates, and $\kappa$ is the photon decay rate in resonator. 
\subsection{Instant eigenstate basis} To solve the Lindblad equation (\ref{LindbladEquation}) in the adiabatic (instantaneous eigenstate) basis, we need to transfer our Hamiltonian from the diabatic basis to the adiabatic (instant eigenstate) basis \cite{Suzuki2018}. For this we use the static transfer matrix between bases
\begin{subequations} 
\begin{eqnarray}
	&S&=S^\dagger=\begin{pmatrix}
		\gamma_+ & \gamma_- \\
		\gamma_- & -\gamma_+
	\end{pmatrix}, \label{TrasferMatrixBases}\\
&H^\prime_\text{Q}&=SHS^\dagger=\hbar\frac{\omega_\text{Q}}{2}\sigma_z=\frac{1}{2}\begin{pmatrix}
	\omega_\text{Q} & 0 \\
	0 & -\omega_\text{Q}
\end{pmatrix},
\end{eqnarray}\end{subequations}  where $\gamma_\pm=\frac{1}{\sqrt{2}}\sqrt{1\pm\frac{\varepsilon(t)}{\omega_\text{Q}}}$, and $\omega_\text{Q}=\sqrt{\Delta^2+\varepsilon^2(t)}$ is the instant qubit angular frequency.

Then we solve numerically the Lindblad equation \begin{eqnarray}
	\label{LindbladEquation}
	&\dot{\rho}(t)&\!\!\!\!\!\!\!\!=-\frac{i}{\hbar}\left[H_\text{tot},\rho\right]+\\\nonumber
	&+\frac{1}{2}\sum_{i=1}^3\!\!\!
	&\left[2L_i^\text{D}\rho(t)L_i^{\text{D}\dagger}-\rho(t)L_i^{\text{D}\dagger} L_i^\text{D}-L_i^{\text{D}\dagger} L_i^\text{D}\rho(t)\right],\nonumber \\
\end{eqnarray} where $\rho$ is the density matrix. The theoretical spectroscopic and interferometric results were obtained using the QuTiP \cite{Johansson2012,Johansson2013,Lambert2024} library.

\begin{figure}[ht]
\includegraphics{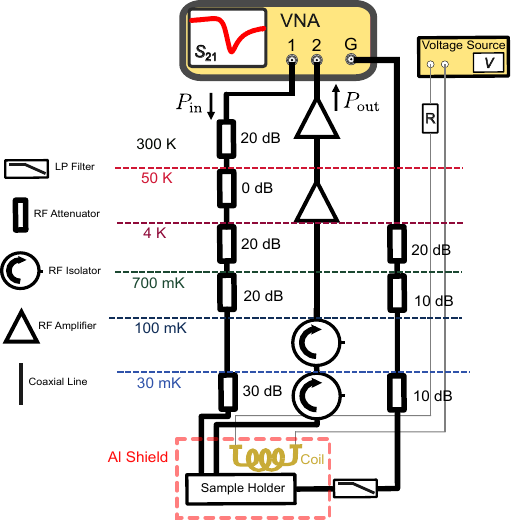}
\caption{Schematic of the experimental setup used in the measurements. Input power is injected through a series of cryogenic attenuators, and the output signal from the sample is amplified through cryogenic and room-temperature amplifiers to obtain $|S_{21}|$. The DC flux point $\Phi_{\textrm{DC}}$ is controlled by a superconducting coil that is current-biased through twisted pair. The AC flux is generated from a microwave generator in the VNA to the on-chip flux bias line.} \label{fig:9} 
\end{figure}

%\label{Sec:AppendixA}
%Appendix 1111
%\subsection{Instant Eigenstate basis}
Consider now qubit-resonator and qubit driving Hamiltonian in the instant eigenstate basis as
\begin{eqnarray}
	%H_\text{Q}^\prime=\frac{\omega_\text{Q}}{2}\sigma_z,\\
	H_\text{C}^\prime=-\hbar g\left(a^\dagger+a\right)\left(\frac{\varepsilon_0}{\omega_\text{Q}}\sigma_z-\frac{\Delta}{\omega_\text{Q}}\sigma_x\right).
\end{eqnarray}
Using the lowering and raising operators for a qubit,
\begin{subequations} 
\begin{eqnarray}
	\sigma=\sigma_+=\frac{1}{2}(\sigma_x+i\sigma_y), \\
	\sigma^\dagger=\sigma_-=\frac{1}{2}(\sigma_x-i\sigma_y),
\end{eqnarray}
\end{subequations} 
that allow us to simplify the interaction and driving operators, we obtain
\begin{eqnarray}
	H^\prime_\text{C}= \hbar\left[g\frac{\Delta}{\omega_\text{Q}}(a^\dagger\sigma+a\sigma^\dagger)+
	g\frac{\Delta}{\omega_\text{Q}}(a\sigma+a^\dagger\sigma^\dagger)-\right. \\ \left. \notag-g\frac{\varepsilon_0}{\omega_\text{Q}}(a^\dagger+a)\sigma_z\right]\approx \hbar g\frac{\Delta}{\omega_\text{Q}}(a^\dagger\sigma+a\sigma^\dagger).
	\label{coupling_approx}
\end{eqnarray}
Here $ g\frac{\Delta}{\omega_\text{Q}}(a\sigma+a^\dagger\sigma^\dagger)$ and $-g\frac{\varepsilon_0}{\omega_\text{Q}}(a^\dagger+a)\sigma_z$ can be neglected in the RWA, because they do not conserve the number of excitations in the system.

Then we can show the diagonalized Hamiltonian for a qubit-resonator system, known as the Jaynes-Cummings Hamiltonian, without driving and probing signals
\begin{subequations} 
\begin{eqnarray}
	H^\prime=H_\text{JC}+H^\prime_\text{D}+H_\text{P} \label{Diagonalized_Hamiltonian}, \\
	H_\text{JC}=H_\text{Q}^\prime+H_\text{R}+H_\text{C}^\prime,\\
	H_\text{Q}^\prime=\hbar \frac{1}{2}\omega_\text{Q}\sigma_z,\\
	H_\text{R}=\hbar \omega_\text{r}\left(a^\dagger a+\frac{1}{2}\right),\\
	H_\text{C}^\prime=\hbar g\left(\sigma_+a+\sigma_-a^\dagger\right).
\end{eqnarray}
\end{subequations} 
This diagonalized Hamiltonian is defined in the same basis as the relaxation and dephasing Lindblad operators.
%That Hamiltonian does not contain all the terms needed for the multi-photon interferometry.

\subsection{Rotating wave approximation, dressed Hamiltonian}

\label{RWA,Dress}
Here we describe a procedure of dressing the Hamiltonian. The dressing technique is convenient to describe systems with two driving signals with significantly different frequencies, like the ones studied here, with probing and driving signals, and it helps to eliminate the fast-rotating terms, which can provide an option to obtain an easier solution for two-tone interferometry \cite{Omelyanchouk2010,Shevchenko2014,Wen2020,Horstig2024}.

Consider our diagonalized Hamiltonian Eq.~\eqref{Diagonalized_Hamiltonian} of the driven system in the rotated basis~\cite{Omelyanchouk2010}
\begin{equation}
	H_\text{RWA}=UH_\text{tot}^\prime U^\dagger+i\hbar\dot{U}U^\dagger,
	\label{DynamicalTransferBasis}
\end{equation}
with the rotating frame with probing frequency $\omega_{\textrm{P}}/2\pi$ as
\begin{equation}
	U=\exp\left[i\omega_\text{P}t\left(a^\dagger a+\frac{\sigma_z}{2}\right)\right],
\end{equation} we then obtain the dressed Hamiltonian in the RWA \eqref{DressedRWAHamiltonian}.

\subsection{Impact of the coupling strength}
We show how the value of the coupling strength impacts  the theoretical interferometry for the transmission coefficient in Fig.~\ref{fig:6} and the qubit upper-level occupation probability in Fig.~\ref{fig:7}.
As we expected, with growing coupling strength, the distortion of the resonances becomes stronger and spreads to higher driving amplitudes. For small coupling the interferometry pattern is very similar to the usual LZSM interferometry \cite{Shevchenko2010,Shevchenko2012,Satanin2012,Ivakhnenko2023}. 
When we increase the coupling to $g=8g_0$ we notice large distortion in resonances, but these calculations may not be fully exact since we neglect counterrotating terms in the coupling in Eq.~\eqref{coupling_approx} and come close to the range of the ultrastrong coupling, where they indeed should be included \cite{Ashhab2010,FriskKockum2019}.

\section{Fabrication methods}
\label{Fabrication_methods}
The fabrication of the device is done in a several  processes on a $675~\mathrm{\mu m}$-thick and highly resistive silicon (Si) substrate, resulting in the device shown in Fig.~\ref{fig:1}(a-c). The fabrication consists of two main steps: (1) fabricating niobium (Nb) structures (resonator, feedline, ground plane, and pads), and (2) fabricating flux qubit made of three junctions of superconductor–insulator–superconductor (SIS) with aluminium (Al) film. A $40~\mathrm{nm}$-thick~AlO$_x$ layer is deposited onto a silicon substrate using atomic layer deposition, followed by a deposition of a $200~\mathrm{nm}$-thick Nb film using DC magnetron sputtering. A positive electron beam resist, AR-P6200.13, is spin-coated with a speed of 6000~rpm for 60~s, and is post-baked for 9 minutes at 160$^{\circ}$C, which is then patterned by electron beam lithography (EBL) and etched by reactive ion etching. A shadow mask defined by EBL on a 1~{$\mathrm{\mu m}$}-thick poly(methyl-metacrylate)/copolymer resist bilayer is used to fabricate the flux qubit made of an Al film connecting the resonator to the ground plane \cite{upadhyay_2021_robust}. Before the deposition of the Al film, the Nb surface is cleaned in-situ by Ar ion plasma milling for 60~s. At the final stage, after liftoff in hot acetone (52 degrees for ~30 minutes) and cleaning with isopropyl alcohol, the substrate is diced  by an automatic dicing-saw machine to the size of 7mm x 7mm and wire-bonded to a copper holder for the low-temperature characterization. The layout of the device and the fabricated sample are shown in Fig.~\ref{fig:8}.

\section{Experimental setup}
\label{Experimental_results}
Measurements are performed in a Bluefors cryogen-free dilution refrigerator with a base temperature $30~\mathrm{mK}$. Using a vector network analyzer (VNA), a probe microwave tone is supplied to the feedline through a $90~\mathrm{dB}$ of attenuation distributed at various temperature stages of the cryostat. The probe signal is then passed through two cryogenic circulators, before being amplified first by a $40~\mathrm{dB}$ cryogenic HEMT amplifier, and secondly by a $40~\mathrm{dB}$ room-temperature amplifier. The dc flux bias is supplied by a nearby superconducting coil connected to an isolated voltage source at room temperature. The drive tone is generated from a microwave generator in the VNA to the on-chip flux bias line through $40~\mathrm{dB}$ attenuation. The device is sealed in a copper holder and covered by an Al shield. The measurement setup is shown in Fig.~\ref{fig:9}.

\nocite{apsrev41Control} 
\newpage
 \bibliography{References}

\end{document}